# On the Trajectory Dependence of Atmospheric Boundary Layer Turbulence Sensing using Small Unmanned Vehicles

**Balaji Jayaraman · Sam Allison · He Bai**

**Abstract** Atmospheric turbulence, especially in the near-surface boundary layer is known to be under-sampled due to the need to capture a wide separation in length and time-scales and limitation in the number of sensors. Over the past decade, the use if Unmanned Aircraft Systems (UAS) technology is approaching ubiquitous proportions for a wide variety of applications, especially with the recent FAA relaxation of flying restrictions. From a geophysical sciences perspective, such technology would allow for sensing of large-scale atmospheric flows, particularly, atmospheric boundary layer (ABL) turbulence, air quality monitoring in urban settings where multitude of small, minimally-invasive and mobile sensors can drastically alter our ability to study such complex phenomena. Currently available observational data of atmospheric boundary layer physics is so sparse and infrequent which significantly limits analysis. With the quantity and resolution of the data that can be measured using a swarm of UAS, three-dimensional reconstruction and deduction of coherent structures in ABL turbulence may be feasible. However, key challenges remain in the form of identifying optimal trajectories to fly the UAS to obtain the relevant quantifications of the turbulence, interpretation of the sensor data from mobile sensors and understanding how representative are the sparse measurements of the overall turbulent boundary layer. This leads to many fundamentally interesting questions that are itemized here: (a) How does UAS trajectory influence sensing and measurements of turbulence? (b) How does ABL turbulence impact UAS trajectory? (c) How to design optimal sensing strategy for canonical turbulence? The key to answering these questions requires the study and understanding of the coupled system of sUAS flight dynamics, controller and ABL turbulence. It is also worth mentioning that some of the above are relevant issues only for small UAS such as quadcopters whose trajectory can be modulated to ABL gusts as against medium-scale fixed wing UAS. In this paper, we leverage a unique sUAS-in-ABL simulation infrastructure that couples high fidelity Large-eddy Simulation (LES) of the ABL with 6-DOF model for the

Balaji Jayaraman, Sam Allison and He Bai
School of Mechanical and Aerospace Engineering, Oklahoma State University
Tel.: +1-405-744-6579
E-mail: balaji.jayaraman@okstate.edu

quadcopter dynamics and a controller for both waypoint navigation and geometric tracking.

# 1 Introduction

One of the major challenges in sensing of atmospheric flows the vast separation of length and time-scales that need to be captured for accurate characterization. This always makes the atmospheric boundary layer or ABL, a highly under sampled system which is especially true when the only source of measurement data was from meteorological towers. Recent advances such the MESONET framework described in Brock et al. (1995) have greatly improved coordinated collection of data, but is till under-resolved. It is this gap that Unmanned Aerial System (UAS) technology , especially the smaller quadcopters, fill by offering the possibility of high resolution measurements. However, unlike fixed met tower type sensing, UAS tend to be mobile with trajectory that in itself is strongly influenced by the field to be measured. The resulting impact on the measured turbulence structure is not well understood and is one of the goals of this study. In addition, questions such as, what is the optimal target trajectory to measure a given turbulent field and the resulting sensitivity of measurements to UAS navigation is not understood either. This is the second major goal driving this work. In order to answer these questions, one way forward is to create realistic sUAS-ABL models that mimic the flight dynamics and navigation performance of sUAS operating in low altitudes is subject to spatially and temporally varying turbulent gusts in the Atmospheric Boundary Layer (ABL). The turbulent flow in the lower atmosphere is a complex dynamical system characterized by the combined influences of the weather, topography, solar heating and diurnal variations, cloud processes, scalar transport and the prevalent state of the atmosphere at the location and time of interest. This interplay produces highly coherent gust patterns that are unique to the conditions of interest Jayaraman and Brasseur (2014), physics-inspired models of sUAS-ABL encounters. While sUAS are associated with $O(1)$m length scales, the most energy containing turbulent motions scale as the distance from the surface at these low-altitude flight trajectories, i.e $O(1-100)$m. In spite of this disparity in length scales, the spatiotemporally varying microscale turbulence impacts the flight dynamics over sub-minute time-scales during which the sUAS traverses potentially hundreds of meters. In this way, the sUAS sees the large0scale variability of the ABL turbulence field. The variability in the velocity across these turbulence eddies can go up to 20% of the mean wind at these low altitudes. Further, in the presence of buoyancy-driven motions in the unstable stability regime, turbulent updrafts are correlated with reduced horizontal wind velocity and the sweeping of these eddies across the sUAS trajectory induces strong space-time variability in the gust patterns that are stability state dependent. These factors make it imperative to consider the larger scale spatial heterogeneity of the turbulence over a wide range of scales. This knowledge can potentially be used to develop navigation strategies to mitigate negative impacts of trajectory deviation of sUAS, for example by improved designs or more intelligent controllers. The other way of leveraging this information is to recover gust information through inverse modeling as is being pursued by the co-authors Allison et al. (2018). From a sensing perspective, modeling these trajectory deviations allows us to realistically estimate the measured turbulent field by the sUAS in its path.

In this study, we will limit our investigations to modeling sUAS dynamics in a canonical horizontally homogeneous ABL operating in quasi-equilibrium over flat topography at a fixed stability state i.e. combination of steady surface heat flux and steady unidirectional "geostrophic" wind vector operating at the mesoscale. To accomplish this, we will use idealized microscale large eddy simulation (LES) to model the extremely high Reynolds number daytime atmospheric boundary layer that is well resolved to accurately capture the energetic atmospheric turbulence motions. The goal is to characterize the space-time variability of the winds relevance to sUAS flight and in turn realistically estimate the measured wind field from an idea sensor mounted on the sUAS. The filtered incompressible Navier-Stokes equation for the resolved velocity field in rotational form along with the Poisson equation for the resolved pressure is used to describe the dynamics of the ABL. The resolved momentum equation contains a sub-filter scale (SFS) stress tensor, Coriolis acceleration and buoyancy force based on the Boussinesq approximation. Virtual potential temperature evolves through a transport equation augmented with a SFS temperature flux vector. The SFS stress in the momentum equation is modeled with a one-equation eddy viscosity model with a prognostic equation for SFS kinetic energy. In such atmospheric flows, the roughness elements are unresolved and the first grid level is in the inertial surface layer. As a result, total surface shear stress is modeled using a locally logarithmic velocity profile. The LES algorithm is highly parallelized, employs spectral difference in the horizontal and finite difference in the vertical and RK3 for the time-integration. The simulation of the quadcopter is accomplished using a 6DOF flight dynamics model along with a PID position controller and a PD attitude controller. This integrated system will be leveraged to model a aUAS ABL sensing paradigm to answer the relevant research question posed earlier.

- Plot the Lumley triangle and invariants
- Plot correlation between deviation form real values and alignment with the mean wind. Plot a 2D polar coordinate diagram with deviation form deviation from mean wind as $\theta$ and the magnutidue of error as the radius.

## 2 Modeling and Simulation Infrastructure

To accomplish the stated goals, we have developed an sUAS-in-ABL simulation infrastructure that is one-way coupled, i.e. the effect of the sUAS is not fed back into the large eddy simulation (LES) model for the ABL. The canonical ABL used to generate the wind model data for this study is modeled as a rough flat wall boundary layer with surface heating from solar radiation, forced by a geostrophic wind in the horizontal plane and solved in the rotational frame of reference fixed to the earths surface. The lower troposphere that limits the ABL to the top is represented with a capping inversion and the mesoscale effects through a forcing geostrophic wind vector. In fact, the planetary boundary layer is different from engineering turbulent boundary layers in three major ways:

1. Coriolis Effect: The rotation of the earth causes the surface to move relative to the fluid in the ABL, which results in angular displacement of the mean wind vector that changes with height.
2. Buoyancydriven turbulence: The diurnal heating of the surface generates buoyancy-driven temperature fluctuations that interact with the near-surface turbulent streaks to produce turbulent motions.
3. Capping Inversion: A layer with strong thermal gradients that caps the microscale turbulence from interacting with the mesoscale weather eddies.

2.1 Large Eddy Simulation (LES) of the Atmospheric Boundary Layer (ABL)

The Reynolds number of the daytime atmospheric boundary layer is extremely high. So, only the most energetic atmospheric turbulence motions are resolved. The passage of the eddies in the surface layer are highly inhomogeneous in the vertical (z), but are clearly homogeneous in the horizontal. In the LES, we try to resolve through the grid as a filter, the energy containing eddy structures which normally account for the majority of the energy content in the turbulent motions. Using a grid filter we can split the fluctuating instantaneous velocity and potential temperature into a resolved and sub-filter scale (SFS) components. The canonical, quasi-stationary equilibrium ABL is driven from above by the horizontal mesoscale 'geostrophic wind' velocity vector, Ug, and the Coriolis force is converted to a driving mean pressure gradient in the horizontal plane perpendicular to Ug. In ABL LES the viscous force is negligible and the surface roughness elements of scale z0 are not resolved by the first grid cell ($z_0 \ll \Delta z$). Buoyancy forces are accurately predicted using the Boussinesq approximation. The momentum equation for resolved velocity contains a sub-filter scale (SFS) stress tensor that is modeled using an eddy viscosity formulation with the velocity scale being generated through a 1-equation formulation for the SFS turbulent kinetic energy. The LES equations are shown below in Eq. (1)-(3) and simulation design is illustrated in Fig.1. The detailed discussion of the numerical methods is available in Khanna and Brasseur (1997); Jayaraman and Brasseur (2014).

$$\nabla \cdot \tilde{u} = 0 \qquad (1)$$

$$\frac{\partial \tilde{u}}{\partial t} + \nabla \cdot (\tilde{u}\tilde{u}) = -\nabla p^* - \nabla \cdot \tau_u^{SFS} + \frac{g}{\theta_0}\left(\tilde{\theta} - \theta_0\right) + f \times (u_g - \tilde{u}) \qquad (2)$$

$$\frac{\partial \tilde{\theta}}{\partial t} + \nabla \cdot \left(\tilde{\theta}\tilde{u}\right) = - - \nabla \cdot \tau_\theta^{SFS} \qquad (3)$$

For this study, we chose a domain size of $400m x 400m x 600$ with the initial capping inversion of height of 280m as shown in fig. 2. The LES grid is designed to be $200x200x300$ for a resolution of 2m in each spatial direction. The surface heat flux is set to zero to realize a neutral ABL and Coriolis parameter is set $0.0001s^{-1}$ (representative of midwest US regions). The roughness height is set to 16cm to mimic grasslands. The geostrophic wind that represents the mesoscale pressure gradient forcing the ABL, $U_g$, is set to 8m/s. The dynamical system is solved using the pseudo-spectral algorithm in the horizontal directions with periodic boundary conditions and second-order finite difference in the vertical using third-order Runge-Kutta time integration. The boundary conditions used in the simulation is as follows: On the upper boundary, the temperature gradient is specified from the initial condition, mean velocity is set as horizontal geostrophic wind vector $U_g$, vertical fluctuating velocity is set to zero and zero gradient is applied to transverse velocity fluctuations and SFS viscosity. On the lower surface the temperature flux is prescribed and normal velocity is zero. In the Poisson solution for pressure, normal pressure gradient is specified using an equation derived from momentum. At t=0, the solution field is initialized using random noise at the first four grid levels near the wall to help accelerate the evolution of the turbulent flow field.

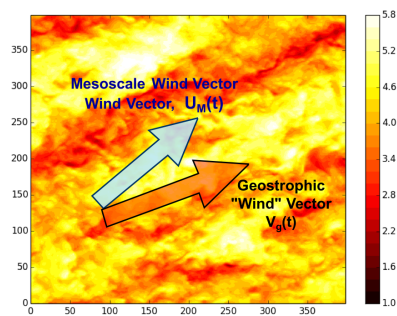

**Fig. 1** Schematic showing the Coriolis effect in a 2D visualization of ABL turbulence for a neutrally stable ABL with $z_i/L = -$ and 10% boundary layer height, $z_i$, using LES.

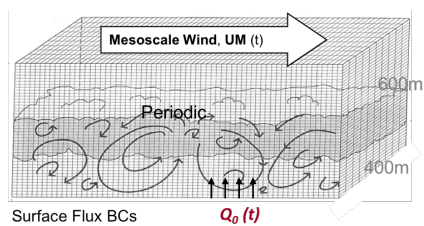

**Fig. 2** Schematic showing the simulation domain with details.

The simulation results are validated in the following way. For a constant value of the mesoscale wind and surface heat flux, the turbulent flow field evolves into a fully developed equilibrium boundary layer. After verifying the existence of statistical stationarity, converged statistics were plotted. A key test to assess the fidelity of the simulations performed is to validate the existence of the Monin-Obukhov (M-O) similarity theory in the near-wall surface layer, which states that, the non-dimensionalized mean wind and temperature statistics under non-neutral conditions are solely a function of the non-dimensional height parameter of stability state parameter, $-z_i/L$. In the above, the non-dimensionalization is performed using the appropriate choice of near-surface parameters for length (distance from the wall) and velocity (friction velocity). In the case of the neutral boundary layer this boils down to the well-known law-of-the-wall. To validate the computational framework and normalized mean velocity gradient is computed for two different ABL simulations corresponding to a neutral and non-neutral stability states as shown below in fig. 3. The other assessment we use to establish the fidelity of the computations is a metric based on the Reynolds-stress tensor. In particular, we compute the inclination of the eigenvector corresponding to the smallest eigenvalue of the tensor with respect to the vertical direction.

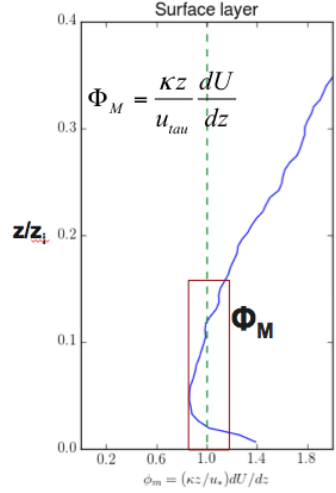

**Fig. 3** Schematic showing the nondimensional mean velocity gradient $\Phi_M$ that should be unity in the logarithmic region of the neutral ABL. Our results show values near unity with small overshoot near the surface which is an artifact in all LES models and surface boundary conditions, also known as the log-layer mismatch.

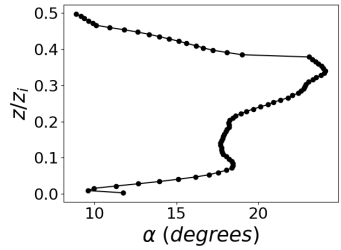

**Fig. 4** Schematic showing the variation of the inclination of the eigenvector corresponding to the smallest eigenvalue of the Reynolds stress tensor with the vertical direction. Literature Klipp (2018) shows that this tilt should be nearly $15 - 20$ degrees near the surface.

The analysis of the data for $\Phi_M$ and $\alpha$ clearly inform us from figs. 3 and 4 about the fidelity of the computations. While the $\Phi_M$ values are predicted in the correct range, the exact match with empirical phenomenology is elusive. It is worth noting that these deviations have more to do with LES design and the overall state of LES turbulence models. As a result, the extent of the agreement shown represents some of the best match with well-accepted empiricism.

2.2 Model for sUAS Dynamics and Control

While there is a standard form for the dynamics of a quadcopter, there are a number of possible controllers that can be used. In this section, we give the formulations for the quadcopter dynamics and the two controllers used.

2.2.1 Dynamics

We will be using the dynamics formulation by Beard Beard (2008) and adding a nonlinear drag term that will be used to apply wind disturbances to the quadcopter

$$\begin{bmatrix} \ddot{p}_n \\ \ddot{p}_e \\ \ddot{p}_d \\ \ddot{\phi} \\ \ddot{\theta} \\ \ddot{\psi} \end{bmatrix} = \begin{bmatrix} (-\cos\phi\sin\theta\cos\psi - \sin\phi\sin\psi)\frac{F}{m} + \frac{f_{d,n}}{m} \\ (-\cos\phi\sin\theta\sin\psi + \sin\phi\cos\psi)\frac{F}{m} + \frac{f_{d,e}}{m} \\ g - (\cos\phi\cos\theta)\frac{F}{m} + \frac{f_{d,d}}{m} \\ \frac{J_y - J_z}{J_x}\dot{\theta}\dot{\psi} + \frac{1}{J_x}\tau_\phi \\ \frac{J_z - J_x}{J_y}\dot{\phi}\dot{\psi} + \frac{1}{J_y}\tau_\theta \\ \frac{J_x - J_y}{J_z}\dot{\phi}\dot{\theta} + \frac{1}{J_z}\tau_\psi \end{bmatrix}. \qquad (4)$$

$(p_n, p_e, p_d)$ are the north, east and down positions of the quadcopter, $\phi$, $\theta$, and $\psi$ are the roll pitch and yaw, $(F, \tau_\phi, \tau_\theta, \tau_\psi)$ are the force and torques due to the rotor thrusts, and $f_d$ is the nonlinear drag force

$$f_d = C_d(V_w - \dot{p})|V_w - \dot{p}|, \qquad (5)$$

where $C_d$ is the drag coefficient matrix and $V_w$ is the wind velocity vector.

2.2.2 Motor and Rotor Models

In addition to the basic dynamics, we include models for the motor dynamics and rotor aerodynamics on the quadcopter. The motor model is based on an experimentally determined brushless DC motor transfer function Xiang et al. (2015),

$$H(s) = \frac{2057342}{s^3 + 1895.s^2 + 13412s + 142834}, \qquad (6)$$

where the input is pulse width modulation and the output is the motor's angular rate. Since the motor is uncontrolled, we added a PID controller that ensures convergence to the desired angular rate within 0.2 seconds. In order to map the motor speed to the forces and torques in (4), we use

$$\begin{bmatrix} F \\ \tau_\phi \\ \tau_\theta \\ \tau_\psi \end{bmatrix} = \begin{bmatrix} k_1 & k_1 & k_1 & k_1 \\ 0 & -Lk_1 & 0 & Lk_1 \\ Lk_1 & 0 & -Lk_1 & 0 \\ -k_2 & k_2 & -k_2 & k_2 \end{bmatrix} \begin{bmatrix} \omega_1^2 \\ \omega_2^2 \\ \omega_3^2 \\ \omega_4^2 \end{bmatrix}, \qquad (7)$$

where $k_1$ and $k_2$ are motor thrust and torque coefficients and $\omega$ is the motor angular rate.

For the rotor aerodynamics, we consider two of the more prominent effects - relative thrust and blade flapping Sydney et al. (2013). Relative thrust is an aerodynamic effect caused by airflow through the rotors. This airflow causes a difference between the expected thrust,

$$T = K_T \omega^2, \tag{8}$$

where $T$ is the thrust and $K_T$ is the thrust coefficient, and the actual thrust. In order to model this effect, we use

$$v_i = \frac{v_h^2}{\sqrt{u^2 + v^2 + (v_i + w)^2}}, \tag{9}$$

$$T_{corrected} = \frac{T v_i}{v_i + w}, \tag{10}$$

where $(u, v, w)$ are the air velocities in the body frame.

Blade flapping is the result of a thrust differential on the advancing and retreating edges of the rotor when the quadcopter has non-zero airspeed. The advancing edge of the blade has a higher thrust than the retreating edge of the blade, which causes the blade to bend slightly more on the advancing edge, resulting in a thrust vector that is at a slight angle with the body frame. This angle, $\alpha$, and the resulting thrust vector can be calculated as

$$\alpha = K_f \sqrt{u^2 + v^2}, \tag{11}$$

$$T_{flapping} = \begin{bmatrix} \frac{u}{\sqrt{u^2+v^2}} \sin\alpha \\ \frac{v}{\sqrt{u^2+v^2}} \sin\alpha \\ \cos\alpha \end{bmatrix} T. \tag{12}$$

2.2.3 Waypoint Navigation Controller

There are two common classes of controllers for quadcopters - waypoint navigation controllers and trajectory tracking controllers. Waypoint navigation controllers are given a desired position and the quadcopter attempts to fly to that position. Trajectory tracking controllers, however, are given a full trajectory - for example, a sinusoidal desired position and its derivative (desired velocity), along with the desired attitude angles. For our waypoint navigation controller, we use a saturated PID controller for waypoint navigation and a feedback linearized PD controller for attitude control. The PID controller is described by

$$\begin{bmatrix} \phi^d \\ \theta^d \\ \psi^d \\ \ddot{p}_d^d \end{bmatrix} = \begin{bmatrix} sat(k_p e_{p_e} + k_d \dot{e}_{p_e} + k_i \int e_{p_e}, \phi_{\max}) \\ sat(k_p e_{p_n} + k_d \dot{e}_{p_n} + k_i \int e_{p_n}, \theta_{\max}) \\ \psi \\ k_p e_{p_d} + k_d \dot{e}_{p_d} + k_i \int e_{p_d} \end{bmatrix}, \tag{13}$$

Atmospheric Boundary Layer Sensing using Unmanned Aerial Vehicles          9where the $d$ superscript designates a desired value, $k_p$, $k_i$, and $k_d$ are the proportional, integral, and derivative gains, $\phi_{\max}$ and $\theta_{\max}$ denote the maximum allowed roll and pitch angles, and the errors are given by

$$\begin{bmatrix} e_{p_e} \\ e_{p_n} \\ e_{p_d} \end{bmatrix} = \begin{bmatrix} p_e^d - p_e \\ p_n^d - p_n \\ p_d^d - p_d \end{bmatrix}. \tag{14}$$

The saturation function, $sat(x, x_{\max})$, is used to ensure that the waypoint navigation controller does not request an unfeasible attitude angle. It is defined as

$$sat(x, x_{\max}) = \begin{cases} -x_{\max} & \text{if } x < -x_{\max} \\ x & \text{if } -x_{\max} \leq x \leq x_{\max} \\ x_{\max} & \text{if } x > x_{\max} \end{cases}. \tag{15}$$

Using the desired roll, pitch, and yaw generated by (13), we control the attitude angles using

$$\begin{bmatrix} F \\ \tau_\phi \\ \tau_\theta \\ \tau_\psi \end{bmatrix} = \begin{bmatrix} \frac{m(g+\ddot{p}_d^d)}{\cos(\phi)\cos(\theta)} \\ J_x(-K_1\dot{\phi} - \frac{J_y-J_z}{J_x}\dot{\theta}\dot{\psi}) + K_{p1}e_1 + K_{d1}\dot{e_1} \\ J_y(-K_2\dot{\theta} - \frac{J_z-J_x}{J_y}\dot{\phi}\dot{\psi}) + K_{p2}e_2 + K_{d2}\dot{e_2} \\ J_z(-K_3\dot{\psi} - \frac{J_x-J_y}{J_z}\dot{\phi}\dot{\theta}) + K_{p3}e_3 + K_{d3}\dot{e_3} \end{bmatrix}, \tag{16}$$

where $K_p$ and $K_d$ are the proportional and derivative gains and the errors are

$$\begin{bmatrix} e_1 \\ e_2 \\ e_3 \end{bmatrix} = \begin{bmatrix} \phi^d - \phi \\ \theta^d - \theta \\ \psi^d - \psi \end{bmatrix}. \tag{17}$$

2.2.4 Geometric Tracking Controller

For our tracking controller, we use the formulation by Lee, Leok, and McClamroch Lee et al. (2010). The controller is a nonlinear tracking controller on SE(3) designed specifically for quadcopters. It is described by

$$F = -(-k_x e_x - k_v e_v - mge_3 + m\ddot{x}_d) \cdot Re_3, \tag{18}$$

$$\tau = -k_R e_R - k_\Omega e_\Omega + \Omega \times J\Omega - J(\hat{\Omega}R^T R_d \Omega_d - R^T R_d \dot{\Omega}_d), \tag{19}$$

where $k_x$, $k_v$, $k_R$, and $k_\Omega$ are the control gains, $R$ is the rotation matrix between the inertial and body frames, $\Omega$ is the angular rate in the body frame $e_x$ is the position error, $e_v$ is the velocity error, $e_R$ is the rotation error, $e_\Omega$ is the angular rate error, and $e_3 = [0,0,1]^T$. Using this controller allows us to command the quadcopter to follow more complex trajectories than we could using the waypoint navigation controller.

**3 Results**

In this section, we present some preliminary results realized over the course of development of the above describe sUAS-in-ABL computational infrastructure. As part of this preliminary effort, we explore three different sUAS trajectories to assess their sensitivity to the sensing outcomes. The first trajectory (T1-H) is a s simple hover mode at a given location. In this case, the sUAS is pushed around by the turbulent gusts and the controller tries restore the original position, thus giving rise to locally random-type disturbances. The second trajectory (T2-TW) chosen is a straight line path with a tail wind and the third (T3-HW) is a straight line path with a head wind. The final trajectory chosen is a typical lawnmover type trajectory (T4-LM) to obtain the best possible coverage of the flow. All these three options are summarized in fig. 5. We simulate all these trajectories at heights of 50m, 100,m and 150m which correspond to $z/z_i \approx 0.15, 0.3$ and $0.45$ respectively.

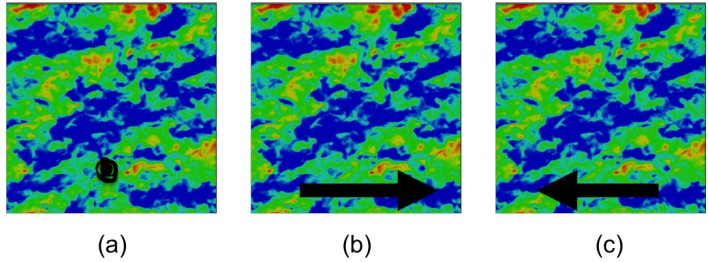

**Fig. 5** Schematic showing the three different trajectories chosen for assessment in this work, namely T1-H (a) for the hover mode, T2-TW (b) for a straight line tail wind mode and T3-HW (c) for the straight line path with head winds.

3.1 sUAS Trajectory Predictions in ABL

The coupled sUAS-in-ABL infrastructure allows us to predict trajectory deviations and quantify their statistical performance. In the schematic shown in fig. 6, the observed deviations in the transverse and vertical directions are $O(1m)$ with stronger deviations in the vertical direction. The reasoning for this is not clear yet and could potentially be due to both the gust information and the controller performance.

3.2 ABL Structure from Different Flight Trajectories

We assume the sUAS is instrumented with a perfect sensor that can measure wind samples. Not clear how it can measure fluctuations, but it can measure the full 3D instantaneous wind. Using the time-series data of measurements from the different trajectories we quantify the uncertainty in the estimated measurements in comparison to the 'true' simulated ABL data. A few of these results are summarized in figs. 7 and 8. The results clearly indicate that the sparse-sampled sUAS measurements can capture the mean velocity field correctly, but not the higher order statistics which indicate trajectory dependence. In particular, we observe sustained over prediction of the second order statistics at all the different measurement heights and for the simpler trajectories considered which hints at under-sampling of the data in spite of the long flight durations considered.

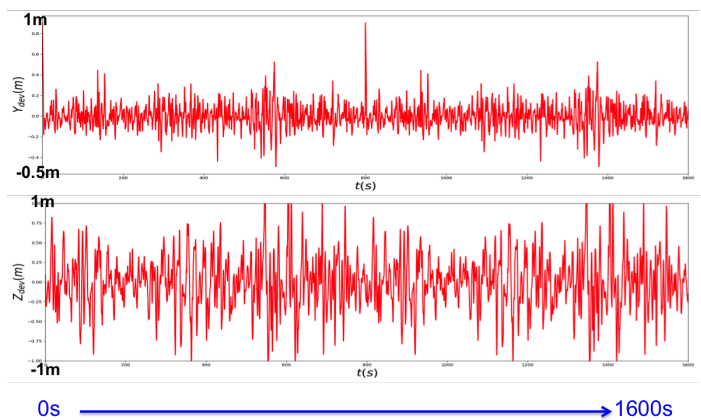

**Fig. 6** Schematic showing the trajectory deviations in the ABL at a height of $z = 50m$ ($z/z_i \approx 0.15$) for the hover mode of the SUAS model. The deviations are $O(1m)$ on average.

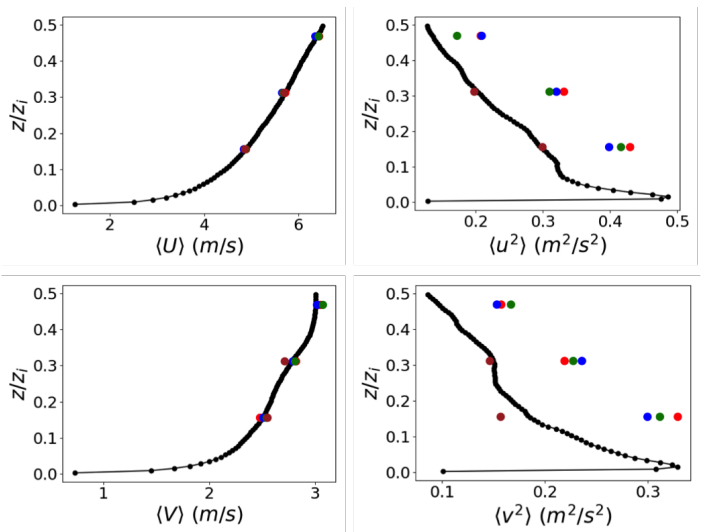

**Fig. 7** Schematic showing the UAS measurements from the different trajectories relative to the data from the entire LES simulation for the duration of the flight. The red dots correspond to T1-H, the blue with T2-TW, the green represents data from T3-HW trajectory and brown corresponds to T4-LM.

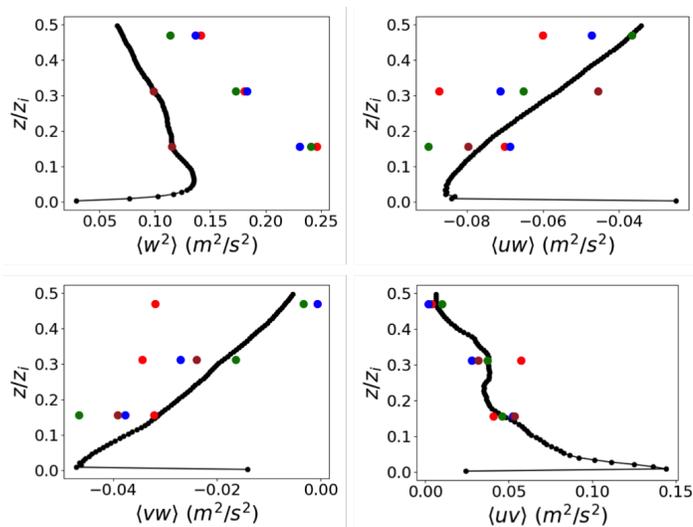

**Fig. 8** Schematic showing the UAS measurements of higher order statistics and covariance measures from the different trajectories relative to the data from the entire LES simulation for the duration of the flight. The red dots correspond to T1-H, the blue with T2-TW, the green represents data from T3-HW trajectory and brown corresponds to T4-LM.

## 4 Conclusions and Planned Future Work

In this preliminary work, we have put together a power infrastructure that can mimic sUAS flight in realistic ABL turbulence that can be leveraged for both improved UAS modeling as well as evaluation & development of sensing strategies. Preliminary conclusions indicate that sUAS measurements are robust in the prediction of the mean, but show trajectory dependence for predicting the higher order statistics accurately. In the future we intend to explore the following.

1. Assess impact of realistic flight plans on sensing accuracy.
2. Identification of coherent structures from time-synchronized sUAS swarms?
3. Assess role of trajectory deviations on ABL sensing vis-a-vis the choice of the trajectory itself.
4. Explore optimal trajectory design and sensor placement.

Some or all of the above issues will be addressed systematically in time for the final paper submission.